# Current localisation and redistribution as the basis of discontinuous current controlled negative differential resistance in NbO$_x$

Sanjoy Kumar Nandi[1], Shimul Kanti Nath[1], Assaad El Helou[2], Shuai Li[1], Xinjun Liu[3], Peter E. Raad[2] and Robert G. Elliman[1]

Dr. S. K. Nandi, S. K. Nath, A. E. Helou, Dr. S. Li, A/Prof. X. Liu, Prof. P. E. Raad, Prof. R. G. Elliman

[1]Department of Electronic Materials Engineering, Research School of Physics, The Australian National University, Canberra ACT 2601, Australia

[2]Department of Mechanical Engineering, Southern Methodist University, Dallas, Texas, USA

[3]Tianjin Key Laboratory of Low Dimensional Materials Physics and Preparation Technology, Faculty of Science, Tianjin University, Tianjin 300354, China

E-mail: sanjoy.nandi@anu.edu.au (S. K. Nandi)
E-mail: rob.elliman@anu.edu.au (R. G. Elliman)

## Abstract

Devices exploiting negative differential resistance (NDR) are of particular interest for analogue computing applications, including oscillator-based neural networks[1]. These devices typically exploit the continuous S-shaped current-voltage characteristic produced by materials with a strong temperature dependent electrical conductivity[2] but recent studies have also highlighted the existence of a second, discontinuous (snap-back) characteristic that has the potential to provide additional functionality. The development of devices based on this characteristic is currently limited by uncertainty over the underlying physical mechanism, which remains the subject of active debate[3-4]. Here, we use in-situ thermo-reflectance imaging and a simple model to finally resolve this issue. Specifically, we show that the snap-back response is a direct consequence of current localization and redistribution within the oxide film, and confirm the veracity of the model by experimentally verifying predicted material and device dependencies. These results conclusively demonstrate that the snap-back characteristic is a generic response of materials with a strong temperature dependent conductivity and therefore has the same physical origin as the S-type characteristic. This is a significant outcome that resolves a long-standing controversy and provides a solid foundation for engineering functional devices with specific NDR characteristics.

## 1. Introduction

Current-controlled negative differential resistance (NDR) is observed in a wide range of materials including amorphous chalcogenides alloys[5-6], mixed ionic-electron conduction devices[7], and amorphous transition metal oxides (e.g. TiO$_x$ [8], TaO$_x$ [9-11], VO$_x$ [12] and NbO$_x$ [13-16]) and is being used to fabricate devices for brain-inspired computing, including: trigger comparators [17], self-sustained and chaotic oscillators [18-19], threshold logic devices [20-21] and the emulation of biological neuronal dynamics [22-23]. In their simplest form such devices consist of simple metal-oxide-metal (MOM) structures and exhibit a smooth transition from positive to negative differential resistance under current-controlled operation (hereafter referred to as S-type NDR) due to a rapid increase in device conductance. In general this can arise from electronic, thermal or a combination of electronic and thermal processes but for amorphous transition metal oxides it is well explained by a thermally induced conductivity change mediated by local Joule heating [2]. Significantly, this can occur as the result of uniform conduction in the oxide film or filamentary conduction resulting from an electroforming process [3] [4].

Several oxides (e.g. $NbO_x$ [3], $VO_x$ [27], $TaO_x$ [9], $NiO_x$ [28], and $SiO_x$ [29]) have also been observed to exhibit a second discontinuous or "snap-back" characteristic (hereafter snap-back) that is manifest as an abrupt, hysteretic voltage change during bidirectional current scans [9, 29-30]. Devices exhibiting more complex combinations of S-type and snap-back characteristics have also been reported and have the potential to afford new functionality [31]. However, while the origin of the S-type characteristics is generally agreed, the origin of the snap-back response remains controversial [4]. For example, Kumar et al. [3] used in-situ temperature mapping to investigate both S-type and snap-back NDR in $NbO_x$-based devices and found that the onset temperature for the former was ~400 K, consistent with a conductivity change due to Poole-Frenkel conduction, while that of the latter was ~1000 K, close to the known Mott-Peierls insulator-metal transition (IMT) in $t-NbO_2$ [32-34]. On this basis, and having observed the $t-NbO_2$ phase in related devices, it was concluded that the snap-back response was a direct result of this transition. However, the IMT based mechanism lacks the generality to account for similar snap-back behaviour observed in oxides such as $TaO_x$ [9], $NiO_x$ [28], and $SiO_x$ [29]. It has also been questioned by Goodwill et. al. [4] based on finite-element modelling of $TaO_x$ and $VO_x$ devices that reproduces the snap-back response without recourse to a material specific phase transition. This modelling further showed that the snap-back response was correlated with redistribution of the current distribution into regions of low and high current density.

We have recently developed a model that explains a diverse range of observed negative differential resistance characteristics, including the snap-back response [31]. This was achieved by explicitly accounting for a non-uniform current distribution in the oxide film and its impact on the effective circuit of the device. In general, the resulting current voltage characteristics can be determined from finite element modelling of the current distribution and the feedback created by local Joule heating. However, for strongly localised distributions the essential physics is well demonstrated by a simple two-zone, lumped-element model in which the current distribution between top (TE) and bottom (BE) electrodes is approximated by a high current-density core represented by an archetype memristor (threshold switch) and a low current-density shell represented by a resistor, as shown in Figure 1a. The NDR characteristics then depend on the relative magnitudes of the shell resistance ($R_S$) and the maximum negative differential resistance of the core ($R_{NDR}$), with S-type characteristics predicted for $R_S > R_{NDR}$ and snap-back characteristics predicted for $R_S < R_{NDR}$ as depicted in Figure 1b,c (see supplementary information). Since the magnitude of $R_S$ depends on the conductivity ($\sigma_{oxide}$), area (*A*), thickness (*t*) and temperature (T) of the oxide film, the model predicts that the transition from S-type to snap-back characteristics can be directly controlled by these parameters.

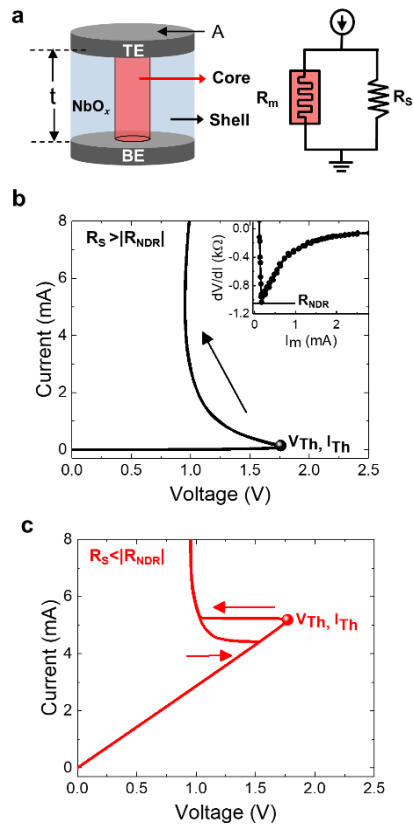

**Figure 1: The core-shell model of memristor and switching characteristics. a)** Schematic of the core-shell representation of the current distribution, and an equivalent lumped-element circuit model of the core-shell structure, where $R_m$ represents a memristive element that reflects the temperature-dependent conductivity of the core region and $R_S$ is the parallel shell resistor. TE and BE refer to top-electrode and bottom electrode, respectively. **b)** Simulated current-voltage characteristics showing continuous S-type NDR with the inset showing maximum negative differential resistance (~1030 Ω), and **c)** Simulated current-voltage characteristics showing discontinuous ("snap-back") NDR behaviour for a parallel shell resistance of 350 Ω ($R_{NDR} > R_S$). Details of the modelling are provided in the Supplementary Information.

Here, we use in-situ thermo-reflectance imaging and quasi-static current-voltage characteristics of NbO$_x$-based cross-point devices to determine the origin of the snap-back mode of NDR and validate the assumptions and predictions of the core-shell model of NDR. This is achieved by correlating NDR characteristics with device current distributions and by demonstrating that NDR characteristics transition between S-type to snap-back modes at critical values of film conductivity, area, thickness and temperature as predicted by the core-shell model.

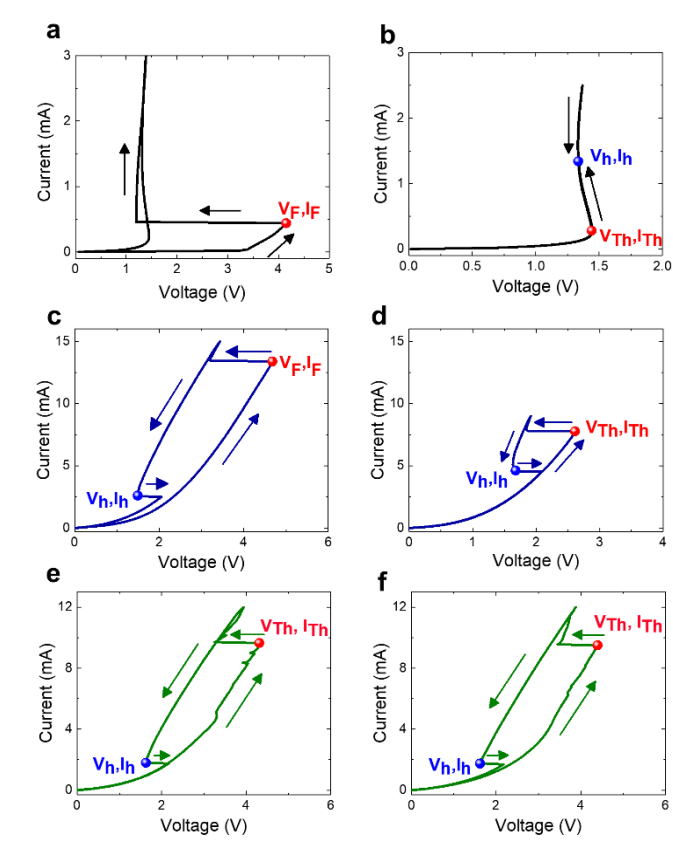

**Figure 2**: **Electroforming and NDR characteristics as a function of stoichiometry**. **a)** Electroforming and **b)** subsequent S-type NDR characteristics of a 10 μm × 10 μm cross-point device of with 25 nm Pt/5nm Nb/NbO$_{2.6}$/25 nm Pt structure. **c)** Electroforming and **d)** subsequent snap-back NDR characteristics of a 10 μm × 10 μm cross-point device with 25 nm Pt/5 nm Nb/44 NbO$_{1.92}$/25 nm Pt structure. **e)** Electroforming and **f)** subsequent switching snap-back characteristics a 10 μm × 10 μm cross-point device with 25 nm Pt/5 nm Nb/44 NbO$_{1.92}$/25 nm Pt structure indicating switching due to current bifurcation. Note that switching characteristics of all samples were measured with negative bias applied to the top electrode unless otherwise stated.

## 2. Results and Discussions

### 2.1 Effect of film conductivity on NDR modes

The effect of oxide conductivity on electroforming and NDR characteristics was assessed using 10 μm × 10 μm cross-point devices with low (x=2.6) and high (x=1.92) conductivity $NbO_x$ films. (The relationship between composition and conductivity is included in the Supplementary Information). Electroforming was conducted in current-control mode using bi-direction current sweeps, with typical results for the low conductivity film shown in Figure 2a. This produces a sudden reduction in voltage as the current is increased beyond a critical threshold value and produces a permanent change in resistance, from a few MΩ before forming to ~10-20 kΩ after forming, consistent with the creation of a permanent conductive filament. After forming, the devices exhibited stable S-type characteristics as shown in Figure 2b. In contrast, devices with high conductivity films exhibited a snapback-characteristic during electroforming, as shown in Figure 2c. In this case electroforming had only a minor impact on the device resistance, reducing it from ~5 kΩ to ~3 kΩ, and subsequent current-scans produced a similar snap-back response but with a lower threshold voltage and threshold current, as shown in Figure 2d.

The forming characteristics of the high-conductivity films were also found to depend on the maximum current employed for forming, with devices subjected to lower currents exhibiting snap-back characteristics without the creation of a permanent filament. This is illustrated in Figure 2e-f, which show successive current-sweeps for a high-conductivity film formed with a maximum current of 12 mA, compared to the 15 mA employed for the previous measurement. No significant change in device resistance was observed after the initial snap-back response and identical characteristics were measured during a second sweep, consistent with the current bifurcating into domains of high and low current-density. However, repeated cycling did eventually cause a reduction in resistance and a concomitant reduction in both threshold-voltage and threshold-current, consistent with the creation of a permanent filament [35].

These results are consistent with the predictions of the core-shell model in that the low-conductivity film (x=2.6) (i.e. higher shell resistance), exhibits S-type NDR characteristics, while the high conductivity film (x=1.92) (i.e. low shell resistance), exhibits snap-back characteristics, as expected. Significantly, they also demonstrate that the snap-back response does not depend on the existence of a permanent conductive filament.

## 2.2 In-situ temperature mapping

To further understand the significance of current localisation and the role of the shell region as a current divider, in-situ thermo-reflectance measurements were performed on devices exhibiting S-type and snap-back NDR characteristics. These measurements were performed on NbO$_x$ films with a conductivity intermediate between those above (i.e. x=2.05), and included devices with and without permanent filaments. The results are summarised in Figure 3 which shows measured current-voltage characteristics and temperature distributions for three devices: a post-formed 5 µm device with a permanent filament that exhibits S-type NDR (Figs. 3a-c); a post-formed 10 µm device with a permanent filament that exhibits snap-back NDR (Figure 3d-f); and a device without a permanent filament that exhibits snap-back NDR (Figure 3g-i). The devices with permanent filaments have highly localised temperature distributions over the full range of operating currents, as expected for filamentary conduction, but the response of the surrounding shell temperature is distinctly different in the two devices. For the device exhibiting S-type NDR (Figure 3a) both the core temperature and that of the surroundings shell increase monotonically with current, while for the device exhibiting snap-back NDR (Figure 3d) the temperature of the core increases rapidly during snap-back while that of the surrounding shell decreases. As the temperature distribution reflects that of the current, the reduction in shell temperature is consistent with the role of the shell as a current divider, with the shell current decreasing due to the increase in core conductivity. In contrast to the devices with permanent filaments, the third device has a spatially uniform current distribution in the sub-threshold current range but undergoes bifurcation into high and low current density domains at, or near, the threshold current (Figures 3g-i). As for the filamentary snap-back device this is also associated with a concomitant reduction in shell temperature. These results clearly demonstrate that the snap-back response is associated with current localisation, either due to a pre-existing permanent filament or current bifurcation, and the redistribution of current between low and high current-density domains.

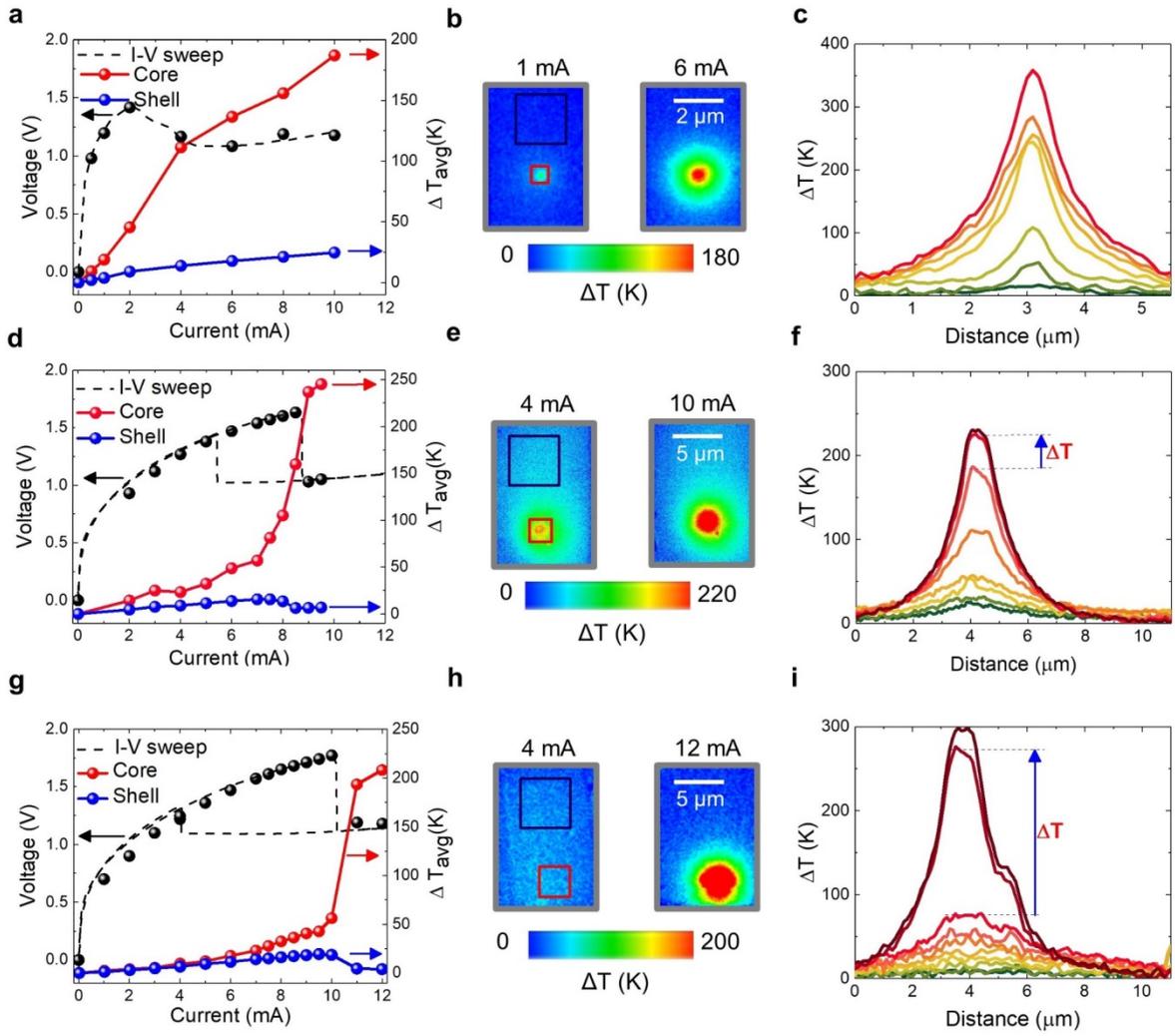

**Fig 3**: **In-situ temperature measurements of S-type and snap-back NDR. a)** Current-voltage (I-V) characteristics and average temperature rise of the filament and surrounding area (the area used for averaging is indicated by boxes in **b**) as a function of applied current, **b)** 2D maps of the surface temperature rise in a 5 µm device operating at 1 mA and 6 mA, **c)** Temperature (current) localisation of S-type NDR in the post-formed device at different current levels as shown in **a**, **d)** I-V characteristics and average temperature rise of the permanent filament and surrounding area (indicated by boxes in **e**) as a function of applied current, **e)** 2D map of the surface temperature rise in a 10 µm device at pre-threshold (4mA) and post-threshold (10 mA) currents, **f)** Temperature (current) localisation in a post-formed device at different currents as shown in **d** (the blue arrow indicates snap-back transition), **g)** I-V characteristics and average temperature rise of the current filament and surrounding area (indicated by boxes in **h**) as a function of applied current, **h)** 2D map of the surface temperature in a 10 µm device at pre-threshold (4 mA) and post-threshold (12 mA) currents, **i)** Temperature (current) localisation due to current bifurcation in the device without permanent filament as shown in **g** (the blue arrow indicates snap-back transition). The device structure used for thermoreflectance measurements was 25 nm Au/5 nm Nb/35 nm $NbO_{2.05}$/40 nm Pt. The circles overlaying the I-V curves represent in-situ I-V measurements during thermoreflectance measurements.

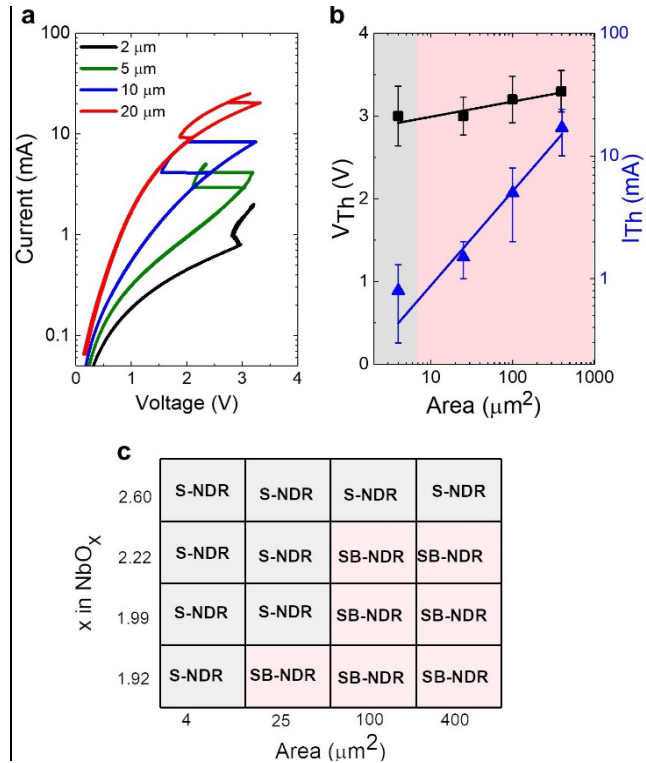

**Figure 4**: **Area dependent switching properties of NbO$_x$ with x=1.92**. **a)** NDR characteristics as function of area showing area dependent S-type NDR and snap-back NDR, **b)** Threshold voltage and current as a function of area. Devices with 25 nm Pt/5 nm Nb/44 NbO$_{1.92}$/25 nm Pt were used to study the area dependent switching. **c)** Matrix representation of dependency of S-type and snap-back NDR as a function of stoichiometry and area. The shaded region in **b,c**, identify devices that exhibit S-type (S-NDR) and snap-back (SB-NDR).

*2.3 Area dependence of NDR modes*

The current-voltage characteristics of Figure 3 also highlight the role of device area in controlling the NDR mode, with the smaller area device (higher shell resistance) exhibiting S-type NDR and the larger area devices (lower shell resistance) exhibiting snap-back NDR. This dependence was investigated more explicitly by comparing the response of devices with different stoichiometry and side lengths of 2 μm, 5 μm, 10 μm and 20 μm. Low-conductivity films (x=2.6) were found to exhibit S-type NDR characteristics independent of device area, demonstrating that the shell resistance satisfied the condition $R_S > R_{NDR}$ for areas as small as 4 um$^2$. In contrast, high conductivity films (x=1.92) were found to have area-dependent forming and NDR characteristics, included an increase in forming current with increasing device area (see Figure S3, supplementary information) and an area-dependent transition from snap-back to S-type NDR characteristics, as shown in Figure 4a. The threshold voltage ($V_{Th}$) for these devices was 3.0±0.4 V and is independent of device area, as shown in Figure 4b. However, the corresponding threshold current (or threshold power, $P_{Th} = I_{Th} \times V_{Th}$) was found to increase monotonically with increasing device area, as expected from the reduced film resistivity. This contrasts with situation for the low conductivity films where the threshold power remains constant with device area[24]. Results for all film compositions are summarised in Figure 4c and reflect the predicted area dependence, with large area devices having lower shell resistance and exhibiting snap-back characteristics, and smaller area devices having higher shell-resistance and exhibiting S-type NDR characteristics.

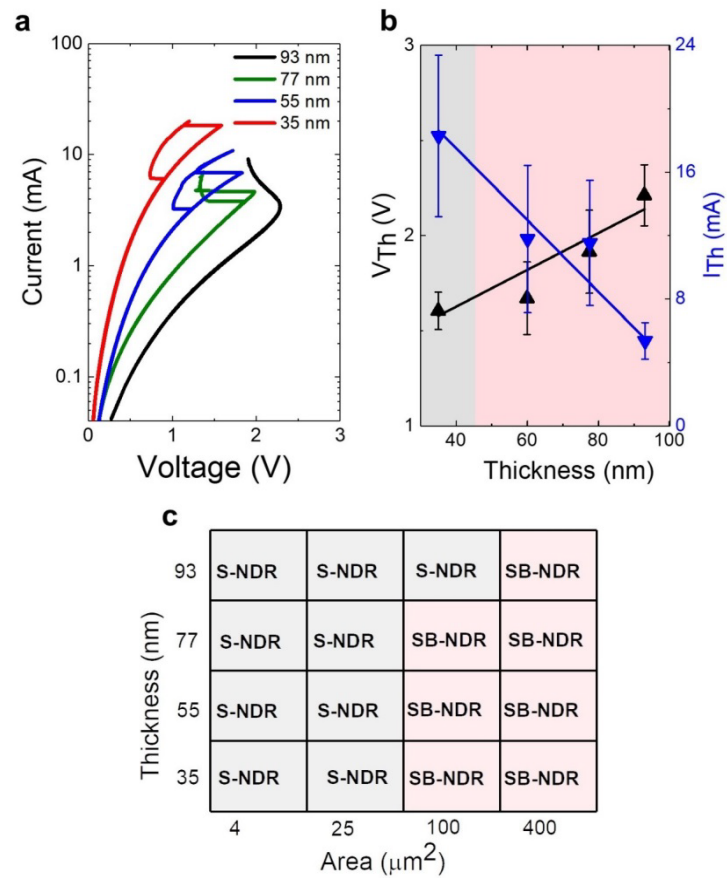

**Figure 5**: **Thickness dependence switching properties of NbO$_x$ with x=2.05**. **a)** Thickness dependent current-voltage characteristics of 10 μm× 10 μm device showing area dependent S-type NDR and snap-back (SB-NDR). **b)** Threshold voltage and current as a function thickness with switching mode depicted by shading. Data points represent averages for ten devices of each thickness. Devices with 25 nm Au/5 nm Nb/NbO$_{2.05}$/40 nm Pt used to study the thickness dependent switching. **c)** Matrix representation of dependency of S-type and snap-back NDR as a function of thickness. The shaded region in **b,c**, identify devices that exhibit S-type (S-NDR) and snap-back (SB-NDR).

## 2.4 Thickness dependence of NDR modes

The effect of film thickness was investigated using devices with dimensions of 10 μm×10 μm and NbO$_x$ films with x=2.05. All devices underwent an initial electroforming step prior to measuring NDR characteristics and the forming current was found to decrease linearly with increasing film thickness, from 33.4±5.2 mA for a film of 35 nm thickness to 10±2.1 mA for a film of 93 nm thickness, consistent with the expected increase in film resistance i.e. $R_{Device} \propto t$ (Figure S3, supplementary information). The subsequent NDR characteristics were then found to depend on film thickness, with films of thickness ≤ 77 nm exhibiting snap-back characteristics and a film of thickness 93 nm exhibiting S-type NDR characteristics, as shown in Figure 5a. This is accompanied by an increase in threshold voltage and a decrease in threshold current (see Figure 5b), consistent with the increase in film resistance with increasing thickness. As for the conductivity and area dependencies, these results are consistent with model predictions based on the assumption that the shell resistance transitions from the condition $R_S > R_{NDR}$ for the 93 nm thick film to $R_S < R_{NDR}$ for films of thickness less than 77 nm. Figure 5c summarises results for devices of different areas and shows that the transition between S-type and snap-back NDR was limited to 100 μm² devices for the thickness range investigated.

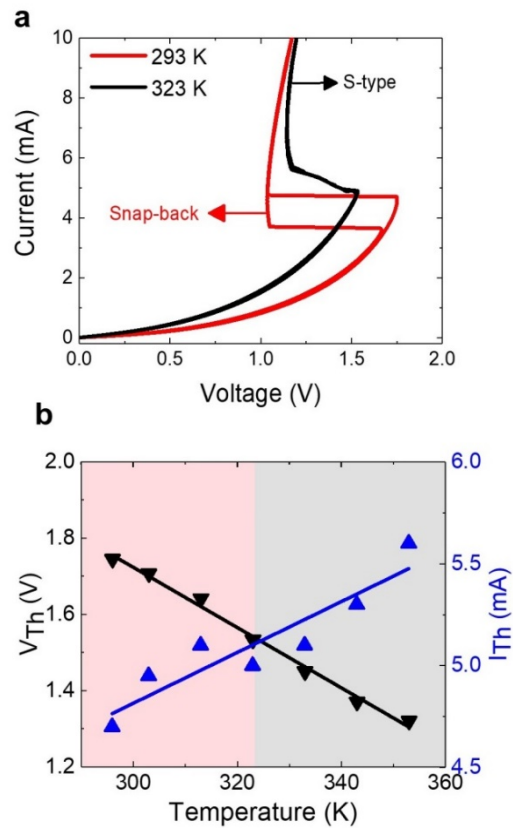

**Figure 6: Effect of substrate temperature on snap-back NDR. a)** Snap-back NDR as a function of temperature showing the transition from snap-back to S-type at ~320 K and **b)** the corresponding threshold -voltage and -current as a function of temperature. The shaded region in **b,** identify devices that exhibit S-type (S-NDR -grey) and snap-back (SB-NDR-red) responses.

*2.5 Temperature dependence of NDR modes*

Finally, we consider the effect of substrate temperature on NDR characteristics. The devices used for these studies had a NbO$_x$ composition of x=2.22. These devices exhibited a snap-back response at 293 K after electroforming but transitioned to an S-type response as the temperature was increased to 323 K, as shown in Figure 6a. This is counterintuitive, as the resistance of the shell is expected to decrease with increasing substrate temperature and to produce a transition from smooth to snap-back characteristics as the temperature exceeds the point where $R_S < R_{NDR}$. The increase in oxide conductivity with temperature is clear evident from the decrease in threshold voltage ($V_{Th}$) and the increase in threshold current ($I_{Th}$) shown in Figure 6b. Given the transition criterion $R_S < R_{NDR}$, these results suggest that the core $R_{NDR}$ also varies with temperature and that it decreases more rapidly with temperature than the surrounding oxide shell. However, this is remains the subject of further studies.

The above results clearly demonstrate that the NDR response of NbO$_x$-based cross-point devices can be controlled by the oxide conductivity, area and thickness and that the snap-back mode of NDR is associated with redistribution of current from low to high current-density domains in a manner consistent with the proposed core-shell model of current transport. Significantly, the results also explain several interesting results reported in the literature and resolve a long-standing controversy about the origin of the snap-back mode of NDR. Specifically, the area dependence of the NDR response observed in the finite element simulations by Goodwill *et al* [4] can now be understood in terms of the relative resistances of the core and shell regions and the associated current redistribution. However, as we have shown this is only one of several dependencies that control the transition from S-type to snap-back NDR. Moreover, the snap-back is not predicated on current bifurcation but simply requires current localisation, as is evident from the response of electroformed devices that retain a permanent filament (Figure 3d-f). In that case, current confinement is preordained and the dominant NDR mode is determined mainly by the resistance of the shell region. Finally, because the snap-back NDR derives from the same physical mechanism as the S-NDR it has a similar onset temperature, estimated to be ~400-600 K (as shown in Figure 3)[,2] As this is simply determined by the temperature dependence of the film conductivity[2] it explains why snap-back NDR is observed in a diverse range of amorphous metal oxides and obviates the need for a material-specific phase transition[3]. It can also explain the origin of S-type characteristics in t-NbO$_2$[36] and crystalline VO$_2$-based devices[37].

## 3. Conclusion

In summary, we have demonstrated that the NDR characteristics of $NbO_x$ cross-point devices can be controlled by materials and device parameters and that these dependencies are consistent with the predictions of a model that can account for a non-uniform current distribution and the redistribution of current from regions of low to high current density. This was achieved by using in-situ thermoreflectance imaging to correlate NDR characteristics with changes in current distribution, and by systematically studying how the transition from S-type to snap-back NDR depended on the conductivity, area, thickness and temperature of the $NbO_x$ film. These results conclusively demonstrated that the snap-back characteristic is a generic response of materials with a strong temperature dependent conductivity and that it has the same physical origin as the S-type characteristic. This is a particularly significant finding as it resolves a long standing controversy and provides a strong foundation for engineering devices with specific NDR characteristics for brain-inspired computing.

**Experimental Methods**

The devices employed in this study were $Nb/NbO_x/Pt$ cross-point structures fabricated using standard photolithographic processes described elsewhere[38]. The bottom electrode, comprising 5 nm Nb (or 10 nm Cr) and 25 nm Pt (or 40 nm Pt) layers, was deposited by e-beam evaporation onto on a 300 nm thermal oxide layer on a Si (100) wafer. $NbO_x$ dielectric layers of variable composition were subsequently deposited using either RF sputtering of a $Nb_2O_5$ target in an Ar ambient or DC sputtering of a Nb target in a variable $O_2/Ar$ ambient. Details of the deposition conditions are given in the supplementary information (Table S1). Grazing incident-angle X-ray diffraction (GIAXRD) of films deposited onto Si substrates confirmed that they were amorphous, while Rutherford backscattering spectrometry (RBS) of films deposited onto vitreous carbon or Si substrates showed that they had compositions in the range from x=2.60±0.05 to 1.92±0.04 (i.e. oxygen rich $Nb_2O_5$ to sub-stoichiometric $NbO_2$). We have previously studied similar films by Reflection Electron Energy Loss Spectroscopy (RHEELS) and shown that they do not contain metallic Nb[39]. To complete the device structure, we have deposited 25 nm Pt (or 25 nm Au) with a 5 nm Nb adhesion layer.

Electrical measurements were performed using an Agilent B1500A semiconductor parameter analyser attached to a Signatone probe station (S-1160). All measurements were executed under atmospheric conditions by applying voltage on the top electrode, while the bottom electrode was grounded. Note that switching characteristics were measured with negative bias applied to the top electrode unless otherwise stated.

In-situ temperature measurements were performed with a TMX T°Imager® transient thermoreflectance (TR) imaging system using a 100x objective lens and a 490 nm illumination wavelength (see supplementary information). For these measurements DC and pulsed electrical signals were supplied by a Keithley 2410 parameter analyser. In these measurements the measured temperature is proportional to the change in reflectance (R) according to the relation $\frac{\Delta R}{R} = \left(\frac{1}{R}\frac{\partial R}{\partial T}\right)\Delta T = C_{TR}\Delta T$, where $\Delta R$ is the change in reflectance, $\Delta T$ is the change in temperature, and $C_{TR}$ is the thermoreflectance co-efficient for the top electrode material. As the thermoreflectance co-efficient of Au ($C_{TR}=5\times10^{-4}$ K$^{-1}$) is much larger than that of Pt ($C_{TR}=-0.31\times10^{-4}$ K$^{-1}$), we performed thermoreflectance measurements on devices with Au top electrode to improve the TR measurement sensitivity. Further details of the thermoreflectance method can be found in references[40-41]. Finite-element modelling of our cross-bar structures has shown that the temperature difference between the top electrode and oxide layer increases with increasing temperature and top electrode thickness[38]. The calculated surface temperature are broadly consistent with thermoreflectance measurements.

**Acknowledgements**

This work was partly funded by the Australian Research Council (ARC) and Varian Semiconductor Equipment/ Applied Materials through an ARC Linkage Project Grant: LP150100693. We would like to acknowledge access to NCRIS facilities at the ACT node of the Australian Nanotechnology Fabrication Facility (ANFF) and the Australian Facility for Advanced ion-implantation Research (AFAiiR), and thank Dr Tom Ratcliff for comments and feedback on the manuscript.

# Supplementary information

**Core-Shell Model – Stability Criteria**

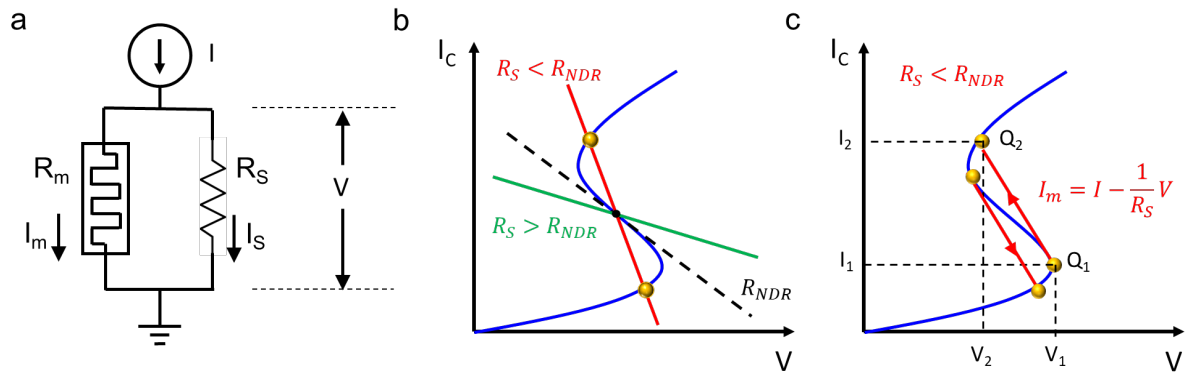

**Figure S1: Conditions for snap-back.** a) Circuit representation of core shell structure, b) I-V characteristics of memristive core showing effect of parallel resistance on the slope of the load-line, c) I-V characteristics for $R_S > R_{NDR}$. (Note: the current-voltage characteristics are in terms of $I_C$ not the total current I.)

Figure S1a shows circuit representation of the core-shell structure, where the core exhibits current-controlled negative differential resistance and has a resistance $R_m = R_m(I_m)$, and the shell has a fixed resistance of $R_S$. The characteristics of the memristive core can be plotted directly on $I_m$ versus V axes, and that of the shell resistor can be included by noting that:

$$I = I_m + I_S \qquad \text{Eq. S1}$$

so:

$$I_m = I - \frac{1}{R_S} V \qquad \text{Eq. S2}$$

Figure S1b shows the I-V characteristic of the memristive core together with a load-line representing the maximum negative differential resistance, $R_{NDR}$. Also shown is the effect of the shell resistance on the slope of the load-line. For $R_S > R_{NDR}$ the intersection between the core and shell characteristics is always single valued, representing stable current controlled NDR. However, for $R_S < R_{NDR}$ the intersection is multivalued, representing a bistable state that gives rise to snap-back.

For a given shell resistance, $R_S$, the point of instability is reached at the threshold current for CC-NDR, at which point $R_C$ becomes negative and a small increase in current leads to an effective reduction in device resistance. The core and shell characteristics are then as shown in Figure S1c, and there is a sudden drop in voltage from $V_1$ to $V_2$ as the current reaches the threshold value, $I_1$. This is accompanied by a sudden increase in the core current from $I_1$ to $I_2$, and a corresponding reduction in shell current from $I_S = I_1 + V_1/R_S$ to $I_S = I_2 + V_2/R_S$, consistent with the redistribution of current observed by thermoreflectance imaging during snap-back.

During the reverse current sweep, the core begins to exhibit CC-NDR at the hold point, so that the snap-back response exhibits hysteresis as shown by the arrows on the $R_S$ load lines.

**Core-Shell Model – Memristive Response**

Lumped element modelling was undertaken using a circuit representation of the memristor core and solved using LT-SPICE as shown in Figure S2 [1]. The active region (core) of the memristor is described by Poole-Frenkel conduction [2] as:

$$R_m = R_0 . exp[(E_a - \sqrt{\beta E})/kT] \qquad \text{Eq. S3}$$

where $R_0$ is the pre-exponential factor and $\beta = \frac{e^3}{\pi \epsilon_0 \epsilon_r}$. The dynamic behaviour of the memristor is described by the lumped element model and Newton's law of cooling as follows:

$$\frac{dT_m}{dt} = \frac{I_m^2 R_m}{C_{th}} - \frac{\lambda_{th} \Delta T}{C_{th}} \qquad \text{Eq. S4}$$

where $\lambda_{th}$ and $C_{th}$ are the thermal conductivity and the thermal capacitance of the device, $\Delta T$ is the temperature difference between $T_m$ and $T_{amb}$. The core-shell structure can be modelled with an archetype memristor as defined above, representing the high-conductivity filament, with resistance $R_m$ defined by Eq.S3, and the parallel conduction through the surrounding film, with resistance $R_S$, as shown in Figure 1(a). Relevant parameters of the model are given in Table S1.

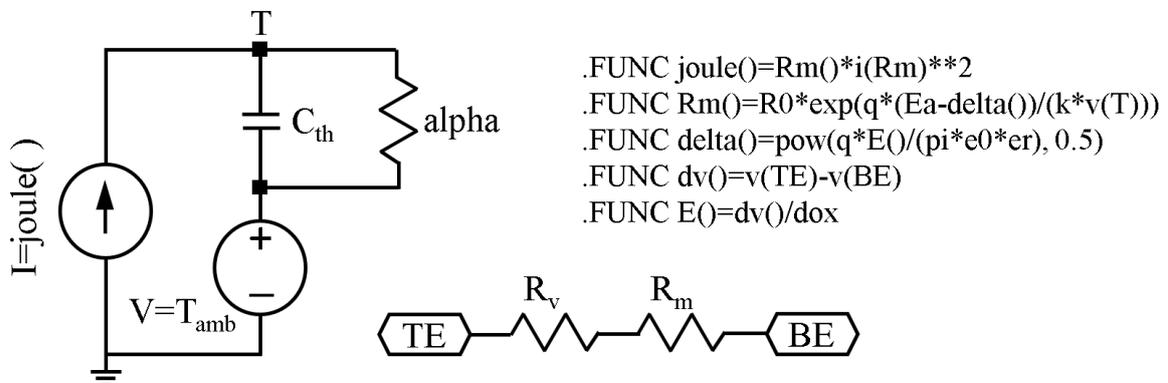

**Figure S2.** SPICE model for threshold switching memristor based on Joule heating and standard Poole-Frenkel conduction [1].

**Table S1.** Memristor parameters used in dynamical simulation based on the Poole-Frenkel conduction model [1].

| Model Parameters (Unit) | Symbol | $R_m$ Value |
|---|---|---|
| Thermal capacitance (J·K$^{-1}$) | $C_{th}$ | $2.5 \times 10^{-13}$ |
| Resistance prefactor (Ω) | $R_0$ | 57 |
| Thermal resistance (K·W$^{-1}$) | α | $1.5 \times 10^5$ |
| Ambient temperature (K) | $T_{amb}$ | 296 |
| Metallic state resistance (Ω) | $R_v$ | 0 |
| Activation Energy (eV) | $E_a$ | 0.255 |
| Boltzmann constant (J·K$^{-1}$) | $k_B$ | $1.38 \times 10^{-23}$ |
| Elementary charge (C) | e | $1.6 \times 10^{-19}$ |
| Vacuum permittivity (F·m$^{-1}$) | $\varepsilon_0$ | $8.85 \times 10^{-12}$ |
| Relative permittivity of the threshold switching volume | $\varepsilon_r$ | 45 |
| Thickness of the threshold switching volume (nm) | $t_v$ | 35 |

## Cross-Point Devoice Structure

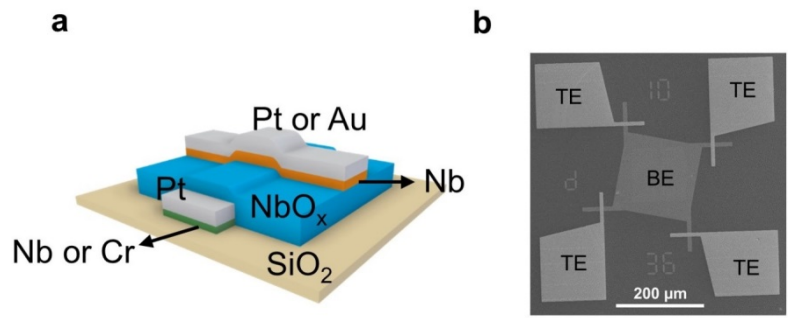

**Figure S3**: **Device structures of the cross-point memristor. a)** Schematic representation of the cross-point device structure with individual layers. **b)** Scanning electron microscopy of 10 μm × 10 μm cross-point devices with a common bottom electrode (BE).

**Table S2**: Device details including the stoichiometry and thickness of each layer.

| Bottom electrode layers | Deposition conditions for NbO$_x$ layers | | | | | | Top electrode layers |
|---|---|---|---|---|---|---|---|
| | Target | Power (W) | Pressure (mTorr) | Gas flow (Ar/O$_2$) | Thickness (nm) | Stoichiometry (NbO$_x$) | |
| 5 nm Nb/25 nm Pt | Nb$_2$O$_5$ | 180 (rf) | 4 | 20/0 | 30 | 2.6±0.05 | 5 nm Nb/25 nm Pt |
| 5 nm Nb/25 nm Pt | Nb$_2$O$_5$ | 180 (rf) | 4 | 20/0 | 45 | 2.6±0.05 | 5 nm Nb/25 nm Pt |
| 5 nm Cr/25 nm Pt | Nb | 100 (dc) | 1.5 | 19/1 | 44.3 | 1.92±0.04 | 5 nm Nb/25 nm Pt |
| 5 nm Cr/25 nm Pt | Nb | 100 (dc) | 2 | 19/1 | 55 | 1.99±0.03 | 5 nm Nb/25 nm Pt |
| 5 nm Cr/25 nm Pt | Nb | 100 (dc) | 1.5 | 18.5/1.5 | 57.3 | 2.22±0.04 | 5 nm Nb/25 nm Pt |
| 10 nm Nb/40 nm Pt | Nb | 100 (dc) | 2 | 18.5/1.5 | 93 | 2.05±0.1 | 5 nm Nb/25 nm Au |
| 10 nm Nb/40 nm Pt | Nb | 100 (dc) | 2 | 18.5/1.5 | 77 | 2.05±0.1 | 5 nm Nb/25 nm Au |
| 10 nm Nb/40 nm Pt | Nb | 100 (dc) | 2 | 18.5/1.5 | 55 | 2.05±0.1 | 5 nm Nb/25 nm Au |
| 10 nm Nb/40 nm Pt | Nb | 100 (dc) | 2 | 18.5/1.5 | 35 | 2.05±0.1 | 5 nm Nb/25 nm Au |

**Film Stoichiometry and Conductivity**

Figure S4a shows a representative RBS spectrum of a sub-stoichiometric NbO$_x$ film deposited on a Si substrate deposited by DC reactive sputter deposition using a mixed Ar/O (19/1) atmosphere (Pressure: 1.5 mTorr, Discharge power: 60 W). The spectrum shows well separated niobium and oxygen peaks that correspond to a stoichiometry of x = 2.2±0.03. Similar analysis was used to determine the stoichiometry of films with x in the range from 1.92±0.03 to 2.6±0.05 (i.e. slightly over stoichiometric). The electrical resistivity of these films was estimated from the low-voltage I-V characteristics of devices with known area and film thickness, and is plotted as a function of film stoichiometry in Figure S4b. These results show that the resistivity has a near exponential dependence on x, decreasing from ~5x10$^3$ $\Omega$.m for x=2.6±0.02 to ~10 $\Omega$.m for x=1.92±0.03.

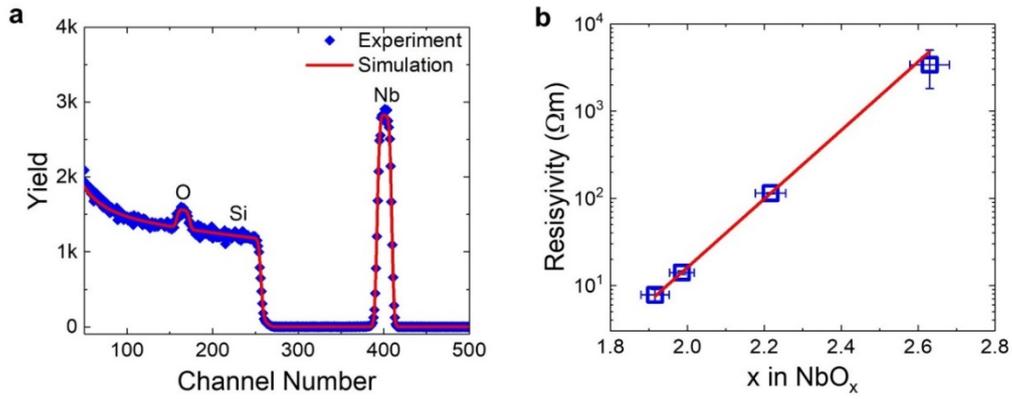

**Figure S4: Effect of stoichiometry on film conductivity. a)** Representative RBS spectrum obtained from a DC reactively sputtered NbO$_x$ thin film deposited on Si substrate at Ar/O (19/1) atmosphere. **b)** Resistivity of the NbO$_x$ film as a function stoichiometry (x).

## Electroforming and NDR Characteristics

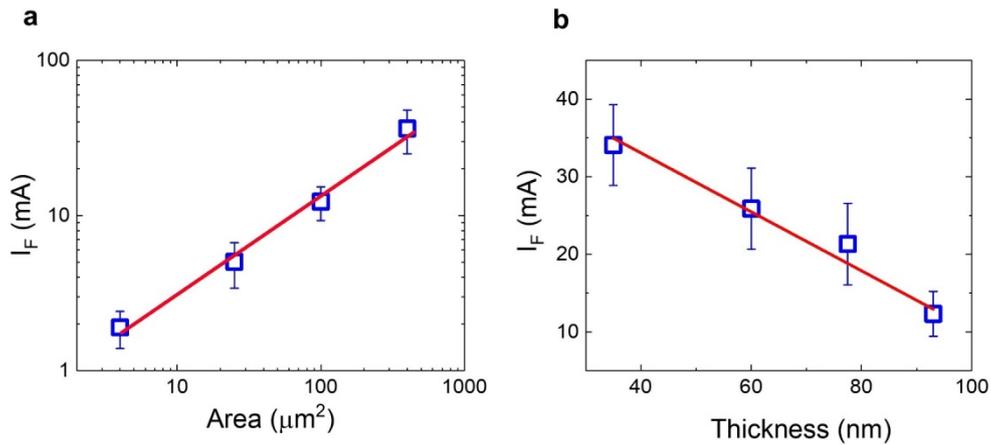

**Figure S5**: **Effect of scaling on electroforming characteristics**. Electroforming current as a function of **a**) area for devices with 25 nm Pt/5 nm Nb/35 NbO$_{1.92}$/25 nm Pt structure and **b**) thickness for 10 μm × 10 μm cross-point devices with 25 nm Pt/5 nm Nb/44 NbO$_{2.05}$/40 nm Pt structure.

20 | P a g e

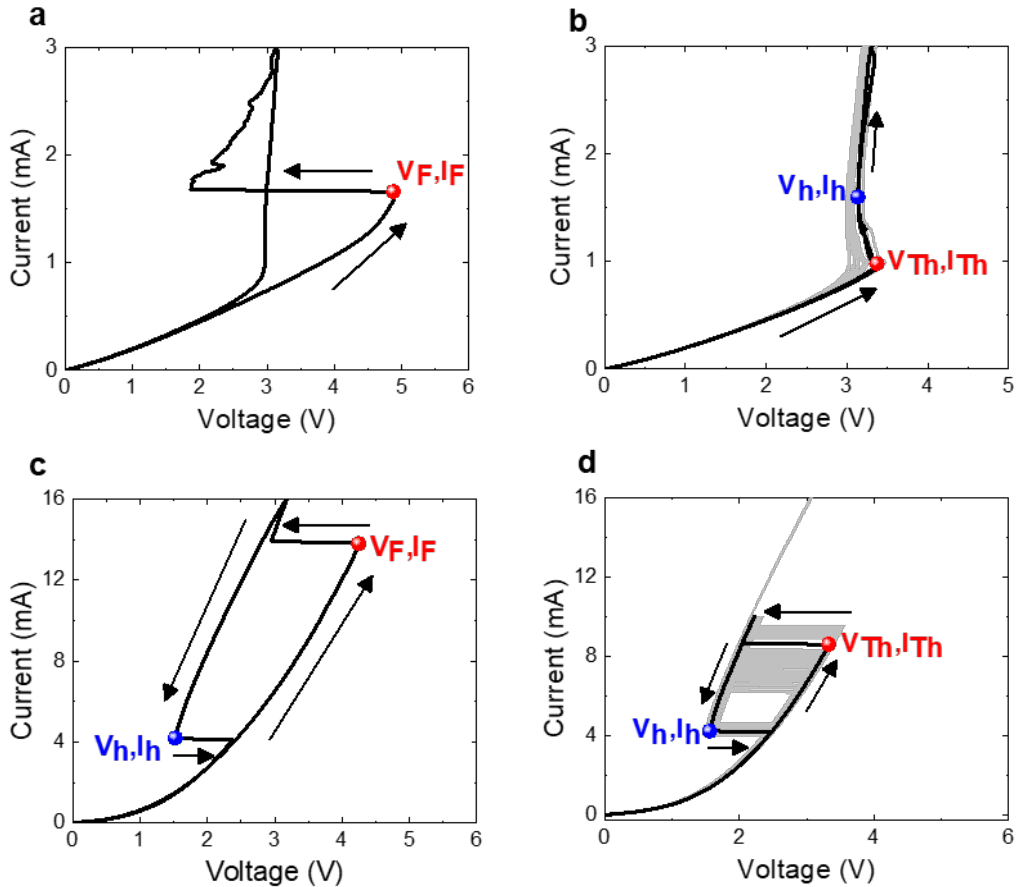

**Figure S6: Electroforming and subsequent NDR characteristics. a)** Electroforming and **b)** subsequent S-type NDR characteristics (10 cycles) of a 2 μm × 2 μm cross-point device. **c)** Electroforming and **d)** subsequent snap-back NDR characteristics (50 cycles) of a 10 μm × 10 μm cross-point device. Both devices have 25 nm Pt/5 nm Nb/44 NbO$_{1.92}$/25 nm Pt structure.

**Thermoreflectance Imaging**

The thermal imaging of the cross-point devices was performed using TMX T°Imager®, a camera-based thermoreflectance imaging system. The thermoreflectance method is based on the temperature dependent optical properties of reflective surface materials, and is non-contact and non-destructive. The method uses visible light illumination, provides deep submicron resolutions (100 nm -300 nm ) well beyond what is possible with other imaging techniques, such as infrared (3 – 10 μm), and is therefore well suited for the wide range of materials present in microelectronic devices.

The temperature-reflectance relation, represented by the Coefficient of Thermoreflectance, $C_{TR}$, describes the unit change of reflectance per degree change in temperature, and is dependent on the surface material, roughness, illumination wavelength, and optical aperture.

For that reason, the $C_{TR}$ coefficient must be obtained in-situ for each device through a calibration procedure where the temperature change is set using a controllable heating stage. During an activation experiment, the change of reflectance intensity due to Joule heating is acquired as the device is activated, and the temperature rise is inferred by combining the reflectance change and $C_{TR}$ fields.

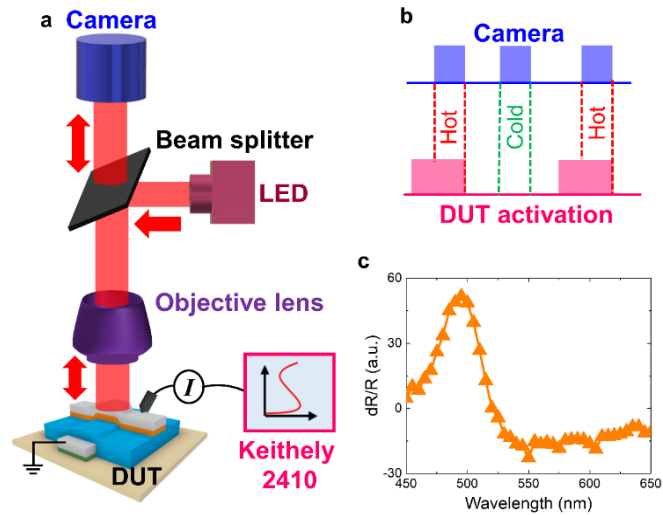

**Figure S7: Thermoreflectance measurement setup. a**) A simple representation of the thermoreflectance measurement system with an objective lenses with a 490 nm illumination, **b**) duration of the hot and cold cycles, where hot refers to the time when a current was applied to the device and cold refers to the time when the device was not activated and **c**, thermoreflectance response of the gold surface under different illumination wavelengths.

The reflectance change is obtained from the change of light intensity collected on a photosensitive detector of a camera for both activation and calibration sequences, providing a 1000×1000 intensity map in the region of interest. For calibration, two temperatures of 20°C and 40°C are set within 0.1°C using a Peltier heating stage, and 50 intensity frames are obtained at each of the two low and high set temperatures. During the measurement, the device is activated with a modulated pulse train while the camera collects 50 "hot" frames during the ON states and 50 "cold" frames during the OFF states. The "hot" and "cold" frames are then used to calculate the unit change in reflectance due to Joule heating during the activation $\frac{\Delta R}{R}(i,j)_{activ.}$, and due to Peltier induced temperature change during the calibration $\frac{\Delta R}{R}(i,j)_{calib.}$.

Equation (1) shows the relations governing the calibration for $C_{TR}$ and the activation temperature relative to the base temperature $\Delta T$ for each pixel $(i,j)$.

$$C_{TR}(i,j) = \frac{1}{\Delta T_{Stage}} \left(\frac{\Delta R}{R}(i,j)\right)_{calib.}$$

$$T(i,j) = T_{base} + \frac{1}{C_{TR}(i,j)} \left(\frac{\Delta R}{R}(i,j)\right)_{activ.}$$

---